\documentclass[a4paper, 10pt]{IEEEtran}  
\usepackage{graphics}
\usepackage{amsmath}
\usepackage{url}
\makeatletter

\usepackage{subfigure}
\usepackage{graphicx}
\usepackage{algorithm}
\usepackage{algorithmic}
\usepackage{times}
\usepackage{verbatim}
\def\unnumfootnote{\xdef\@thefnmark{}\@footnotetext}

\makeatother
\begin{document}
\title{Topological Analysis and Mitigation Strategies for Cascading Failures in Power Grid Networks}
\author{\itshape{S. Pahwa, C. Scoglio, N. Schulz}\\Electrical and Computer Engineering Department\\Kansas State University\\ Manhattan, KS 66506, USA\\email: \{sakship, caterina, noels\}@ksu.edu}
\maketitle
\thispagestyle{empty}
\unnumfootnote{This work is supported by the Energy and Power Affiliates Program, consisting of Westar Energy, Burns and McDonnell, Omaha Public Power District, and Nebraska Public Power District.}
%\unnumfootnote{Electronic addresses:\\sakship@ksu.edu\\caterina@ksu.edu\\noels@ksu.edu}
\unnumfootnote{*This paper is best viewed in colored format}
\begin{abstract}
Recently, there has been a growing concern about the overload status of the power grid networks, and the increasing possibility of cascading failures. Many researchers have studied these networks to provide design guidelines for more robust power grids. Topological analysis is one of the components of system analysis for its robustness. This paper presents a complex systems analysis of power grid networks. First, the cascading effect has been simulated on three well known networks: the IEEE 300 bus test system, the IEEE 118 bus test system, and the WSCC 179 bus equivalent model. To extend the analysis to a larger set of networks, we develop a network generator and generate multiple graphs with characteristics similar to the IEEE test networks but with different topologies. The generated graphs are then compared to the test networks to show the effect of topology in determining their robustness with respect to cascading failures. The generated graphs turn out to be more robust than the test graphs, showing the importance of topology in the robust design of power grids. The second part of this paper concerns the discussion of two novel mitigation strategies for cascading failures: Targeted Load Reduction and Islanding using Distributed Sources. These new mitigation strategies are compared with the Homogeneous Load Reduction strategy. Even though the Homogeneous Load Reduction is simpler to implement, the Targeted Load Reduction is much more effective. Additionally, an algorithm is presented for the partitioning of the network for islanding as an effort towards fault isolation to prevent cascading failures. The results for island formation are better if the sources are well distributed, else the algorithm leads to the formation of superislands. *\\ \\
\end{abstract}
\section{Introduction}

North American Electric Reliability Corporation (NERC) defines a cascading failure as ``The uncontrolled loss of any system facilities or load, whether because of thermal overload, voltage collapse, or loss of synchronism, except those occurring as a result of fault isolation"~\cite{NERC:05}.

In power grids flow dynamics depend greatly on the electrical characteristics such as the voltages, impedances, and the difference in the angles of the voltage phasors of a given pair of buses between which the transmission line is present. If we assume that all the voltages are constant at 1 p.u. and angle difference is very small, we can say that the amount of power flowing through the transmission lines is inversely proportional to their impedances. If a single line gets overloaded or breaks, its power is immediately re-routed to a different line and the disturbance can be suspended. But sometimes, the other line is also already overloaded and must re-route its increased load to its neighbors. This redistribution of power may lead to the subsequent overloading of other lines causing their malfunction at the same time and the consequence could be a cascade of overloading failures.

Such incidents have taken place in history, such as the one on August 10, 1996 when a 1300 MW electrical line in Southern Oregon sagged in summer heat, initiating a chain reaction that cut power to more than four million people in 11 Western States~\cite{BDIA:00},~\cite{MBV:00}. Another example is the incident of August 14, 2003 when an initial disturbance in Ohio~\cite{JR:03} led to the largest blackout in the history of the United States and millions of people throughout parts of  North Eastern and Mid Western United States, and Ontario, Canada, were without power for as long as 15 hours. Electric power systems collapsed in Denmark, Italy, and the United Kingdom within weeks or months of the U.S. blackout~\cite{ISPEC:04}.

Large-scale blackouts are due to the concurrent malfunction of a number of transmission lines and power generators often triggered by the initial failure of a single component of the grid, such as the breakdown of a power line~\cite{PVM:04}. This is discussed in~\cite{JJI:05} with the help of a hidden failure model in which some components of the network such as relays have defects that remain dormant until abnormal operating conditions are reached. The authors of~\cite{JASS:98} have shown that if a line $L$ which shares a node with other lines fails, the hidden failures in all the other lines sharing the node are exposed.

Recently, many complex networks researchers have shown an interest in the topological analysis of real world networks, including power grids as seen in ~\cite{PVM:04},~\cite{MKA:07},~\cite{AY:00},~\cite{BA:02},~\cite{PGVCNA:09},~\cite{ZAR:10},~\cite{RPRV:04}.

While use of advanced control technologies, communication methods and power engineering are critical aspects of providing robustness with respect to cascading failures for power grid networks, a topological analysis is also an important aspect. The power grid can be represented by a large graph belonging to a special family of graphs called complex networks. Power grid networks follow an exponential degree distribution and although not very heterogeneous in the node degree, they show a high heterogeneity in the node load. Most of the nodes handle a small load but there are a few nodes that handle an extremely high load~\cite{PVM:04}. The same is true for links also. Thus, some nodes and links tend to become more important than others and an intentional or accidental removal of these elements can damage the network.
 
Initial studies by~\cite{RHA:00},~\cite{PBCS:02},~\cite{PVMA:03},~\cite{ATY:02} have focused on the static properties of the network, without considering their underlying flow dynamics and have shown that the removal of a node or group of nodes can have important undesirable consequences. Further studies considered flow dynamics based on the efficiency model suggested in~\cite{VM:01}. This model cannot be applied to a power grid directly, and authors of~\cite{PGVCNA:09} agree on this point. They have discussed the problems associated with the general definition of vulnerability based on global efficiency of power grid networks, as mentioned in~\cite{PVM:04},~\cite{AY:00}, and~\cite{VGG:04}. Some modifications to the way in which efficiency is calculated might make it more suitable for the power grid. The authors of~\cite{AY:00} have mentioned that in general, in complex networks, the relevant quantity for the network travels between a set of nodes using the shortest path, and that the load on a node is the total number of shortest paths passing through that node. They have then used this theory to evaluate the robustness of complex networks to cascading failures. Again, this theory cannot be applied directly to power grids unless the concept of shortest paths is precisely defined for power grids. Power does not follow shortest paths based on network metrics such as  number of hops. 

In this paper, we present a thorough analysis on the robustness of power grid networks with respect to cascading failures, calculating flows in the networks using the DC Power Flow Model~\cite{BR:88}. Even if this model is an approximated model where non-linearities are removed, it has been shown to be adequate for our type of analysis. For the purpose of simulating a cascading failure, we developed a simulator and tested it on the IEEE test networks available at~\cite{WAS}. We define Power Degradation as a metric to quantify the load loss in power grid networks, as a result of cascading failure, discussed in Section II. We also generate networks with the same number of nodes and links as the test networks but with a different topology, using a variation of the Generalized Random Graph model for this generation. Comparing the Power Degradation of the different topologies, we show the impact of the topology: among the considered topologies, the best topology provides an improvement of robustness equal to 36.79\%. However, the purpose of this paper is not to generate alternate realistic topologies for the power grid networks, but simply to show that a change in topology can change the robustness of the network to attacks and failures. The provision of realistic guidelines for robust power grid design will be considered in future work.

The second contribution of this paper concerns the definition and testing of two novel mitigation strategies, namely Targeted Load Reduction and Islanding using Distributed Sources. We compare the efficiency of the proposed strategies with respect to the Homogeneous Reduction. Our analysis confirms the efficiency of the Targeted Load Reduction, which reduces on average the load shedding required to stop the cascading effects by $x$\% in total. On the other hand, the Islanding using Distributed Sources is only efficient under some constraints on the level of distribution of the generation. However, when the constraints are met, this strategy can be realistically implemented using small and renewable energy sources such as wind turbines. 

The paper is organized as follows: Section II describes the approach proposed to simulate and analyze cascading failures in a given power grid. In Section III this approach is used to quantify and compare the impact of cascading failures on the  realistic IEEE power grids and the generated graphs. Further, in Section IV, we suggest two mitigation strategies for cascading failures, based on load reduction and network disconnection, with the help of distributed renewable sources. Results and conclusions, along with future work are presented in Sections V and VI, respectively.

Further, in Section IV, we suggest three mitigation strategies for cascading failures - Homogeneous load reduction, Targeted load reduction and Islanding with the help of distributed renewable sources. Among the load reduction strategies, Homogeneous strategy is very simple to implement but Targeted strategy is much more efficient. Islanding is implemented using a two step process to obtain islands which would be powered by distributed energy sources such as wind turbines.
 
\section{Simulation of Cascading Failures}

In this section, we present an approach to analyze the robustness of the power grid network based on the very well-known DC Power Flow Model~\cite{BR:88}, described in details in Appendix $A$. As a result, all the voltages are assumed to be at 1 p.u and the angle differences are assumed to be very small. Thus, the amount of power flowing through each link is approximately inversely  proportional to the impedances. We calculate the 'Maximum power paths' for each node and describe the algorithm for the same in Appendix $B$. Maximum power paths are those which provide the maximum amount of power from a generator to a given node as compared to all other alternate paths from different generators.
The generation and the load are assumed to be equal at all times. Thus, a reduction in load would automatically mean a reduction in the generation. We also make an assumption that the system is at the limit of the maximum load and so the failure of a single component may lead to a cascade of overloading failures.

To analyze the effect of topology of the power grid networks on cascading failures, we used the IEEE 300 bus~\cite{WAS}, IEEE 118 bus~\cite{WAS}, and WSCC 179 bus networks. To explain the power grid as a complex network, the buses are referred to as nodes and the transmission lines as links. Since we consider the DC Power Flow model, we assume that the resistances are very small as compared to the inductances and hence the impedances of the transmission lines are simply reactances. The 300 node network consists of 247 nodes plus two smaller subsections of nodes which are not very well connected with the main graph. These two subsections were not critical for analyzing the power grid from a topological point of view and thus they were not included in the analysis. Therefore, the 300 node network will be referred to as 247 node network here onwards unless we refer to the standard test case. An important attribute to be considered is the capacity of nodes and links. The nodes are characterized by a finite load. The capacity of a node is the maximum load it can handle~\cite{AY:00}. The capacity of a link is the maximum power that it can carry between two nodes. In the case of linearized model, the capacity of a link is primarily governed by the line impedances which depend upon the electrical characteristics of the transmission lines such as length. The amount of power flowing through a line can be calculated using the following relation:
\begin{equation}
P_{ij} = \frac{\delta_{i}-\delta_{j}}{X_{ij}}
\end{equation}
where, $P_{ij}$ is the power flow between nodes $i$ and $j$, $\delta_{i}$ and $\delta_{j}$ are the angles of the voltages at nodes $i$ and $j$, and $X_{ij}$ is the impedance of the link between them.
The angle difference is usually very small. Thus, higher the impedance, lower the capacity.\\
The simulator uses an adjacency matrix $X$ to represent the interconnections in the network. When a link exists between nodes $i$ and $j$, the corresponding entry $x_{ij}$ is a non-zero number and the entry is zero if there is no link present. The non-zero entry in $X$ represents the impedances of the transmission lines. We collected the available load information from the test cases~\cite{WAS} and populated it in a vector called the LoadVector. This vector has a value $0$ if there is no load on the node and a real number if the node carries a load. Nodes can be generators/sources, nodes with no loads, or the load carrying nodes. By Kirchoff's Junction Law, the algebraic sum of incoming power and outgoing power should be zero. Hence, the amount of power that goes into the node should be equal to the sum of the power that is consumed and transmitted further. We assume a lossless situation and therefore no power is dissipated as heat or other losses. As previously mentioned, the total generation is equal to the total load.\\
We categorized the links into two types - vulnerable and non-vulnerable - depending upon whether they caused more or less than ten percent damage to the network upon removal. Damage indicates both, the loss in connectivity and the loss of load. At steady state, approximately $41.97\%$ of the links in the 247 nodes IEEE test network are vulnerable and the remaining $58.03\%$ are non-vulnerable links.
We first calculate the power flowing through each link using the equation above. Then, we select one of the links from the list of vulnerable links and remove it from the network to study the cascading effect. As a result of this removal, the power carried by this link is redistributed among its neighboring links. All the angles and power flows are recomputed by the method described in Appendix $A$. After the initial failure, the power grid goes through multiple stages of failures before it finally stabilizes. Each stage of cascade is called an iteration. We record the number of links and nodes that fail in each iteration as a result of the failures in the previous iteration and the consequent redistribution. We use Power Degradation as a measure to quantify the severity of the damage in terms of load loss. Power Degradation is the current load on the system. In other words, it is the difference between the original load on the system and the load lost in the current iteration. Power degradation is observed at each iteration and the power degradation graph as shown in Fig. \ref{Fig5} is generated. We consider worst case analysis, which means the analysis of the system by removal of a link which causes the maximum damage to the system. Analysis of the graph is done in the following section.

\section{Impact of topology on the robustness of power grid network}
The topology of a network determines the arrangement of nodes in the network and how they are connected to each other. A small change in the network topology such as removal or addition of nodes and/or links may lead to changes in the network properties. This, in turn, may affect the robustness of the network. The aim of this section is not to suggest alternate topologies for a power grid but simply to show that a change in topology can make the power grid more robust to failures and attacks.\\
Since power grids are critical infrastructure, the data of power grid networks is not readily available, except for the few test cases available at the online archive of Washington University~\cite{WAS}. Thus, for a topological comparison, there was a need to generate other topologies which were similar to realistic power grid models but had different arrangement of nodes and links. We keep the number of nodes and links the same as the corresponding test case but connect the links differently as compared to the test case. Since the number of nodes and links of the generated networks is the same as that of the test cases, the average node degree is also the same because average node degree is given by:

\begin{equation}
<k> = \frac{2E}{N}
\end{equation}
where $E$ is the number of links and $N$ is the number of nodes.

\begin{figure}[h]
\centerline{\includegraphics[width = 3in,  height = 2in]{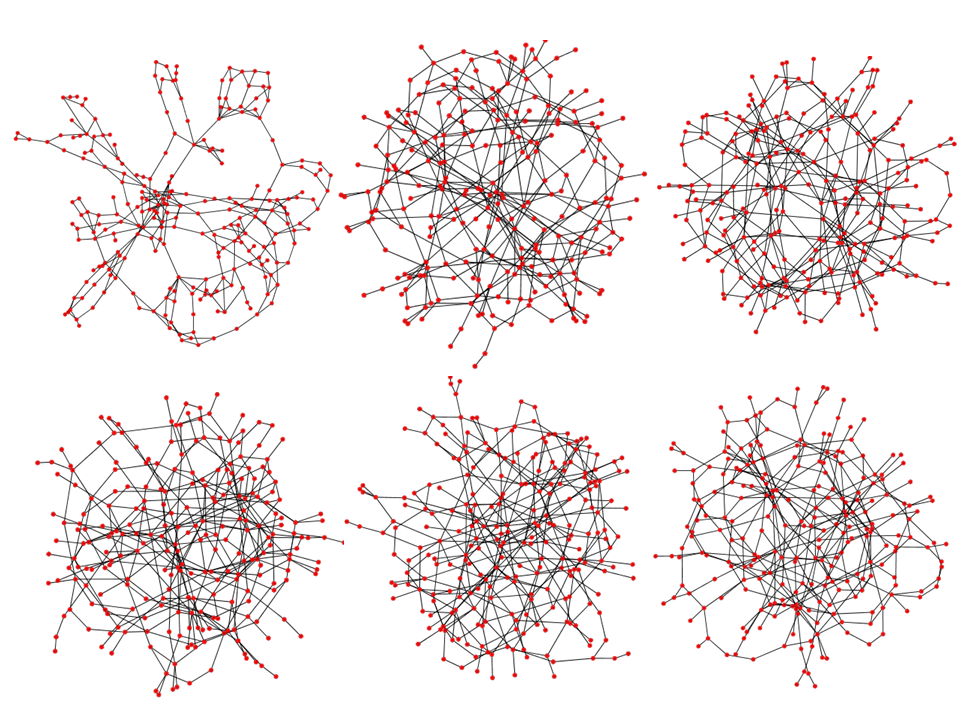}}
	\caption{247 Nodes Original (first) and Generated Networks}
	\label{Fig1}
	\hfill
\end{figure}

\begin{figure}[h]
\centerline{\includegraphics[width = 3in,  height = 2in]{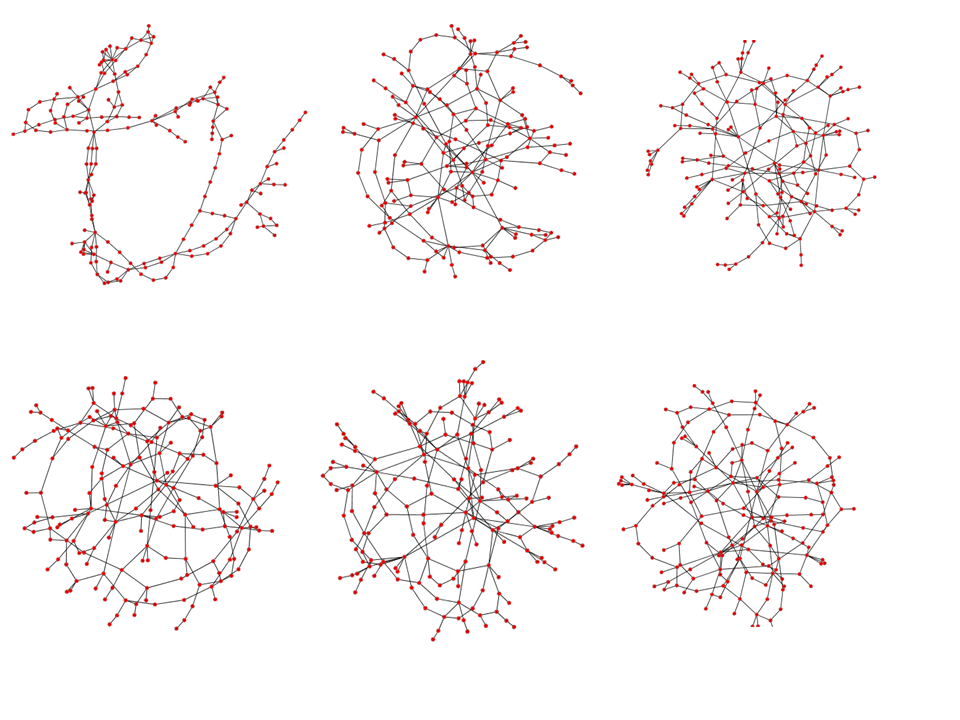}}
	\caption{179 Nodes Original (first) and Generated Networks}
	\label{Fig2}
	\hfill
\end{figure}

\begin{figure}[h]
\centerline{\includegraphics[width = 3in, height = 2in]{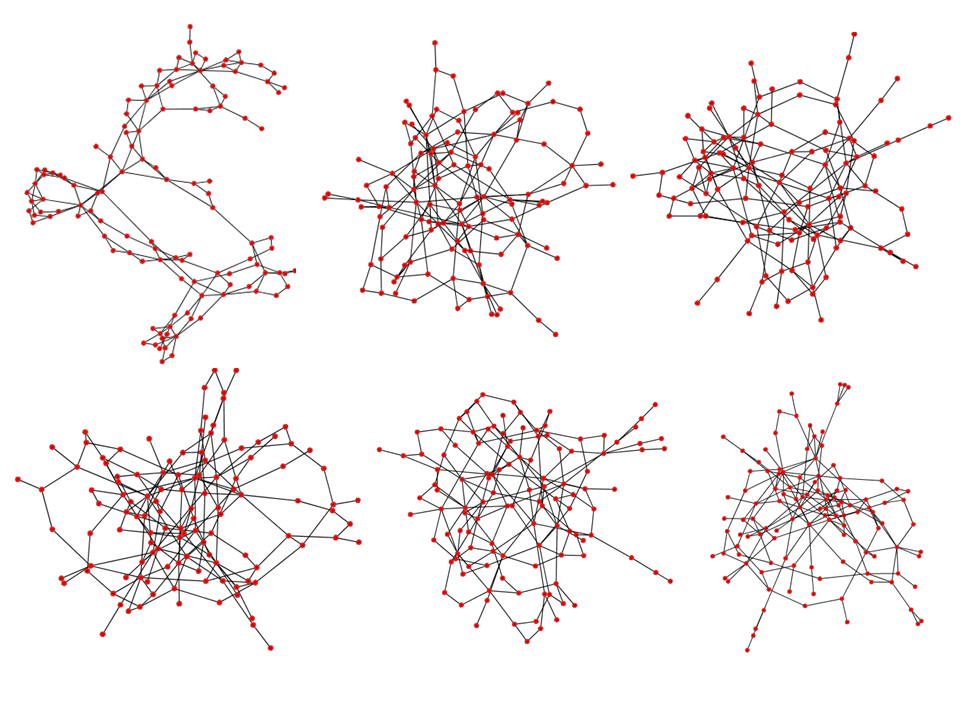}}
	\caption{118 Nodes Original (first) and Generated Networks}
	\label{Fig3}
	\hfill
\end{figure}

In addition, since the generated networks should be as close to realistic networks as possible, we also match the maximum node degree of the generated networks with the corresponding test networks. Since the degree distribution is not kept fixed, we call these generated networks as first approximation networks. We generate about 20 such first approximation networks, out of which 5 are shown in Fig. \ref{Fig1}, Fig. \ref{Fig2} and Fig. \ref{Fig3} for the 247, 179 and 118 node networks respectively.


\subsection{Comparison with the Generalized Model for Random Graphs}
The Erdos-Renyi model~\cite{PA:60} and the network generation algorithm of Molloy and Reed~\cite{MR:98} are the simplest network models to include stochasticity as an essential element in the construction of the network. The connection of links in these two models is completely random and lack knowledge of the principles that guide the creation of links between nodes. Thus, links in these models are created randomly with a given connection probability $p$. Another model for creation of random graphs is the Generalized Random Graph model~\cite{AMA:08},~\cite{MJ:08},~\cite{RVDH:10} which still keeps the assignment of links random but the difference is that it specifies a predefined degree sequence. Thus, the Generalized model can produce graphs with degree distributions which are not necessarily Poisson.
The Generalized Random Graph model was first proposed in 1978 in~\cite{BC:78} and is also known as the Configuration model. This model generates networks with a given degree distribution. It is specified in terms of a degree sequence, for example, for a network with $N$ nodes, there is a fixed degree sequence ${k_{i}}$, for $i = 1 to N$, such that the $ith$ node has degree $k_{i}$. Then two elements are randomly picked from this sequence and the nodes corresponding to those entries are connected. Thus, these graphs are completely random but with an imposed degree distribution.
Our model falls in between the Erdos-Renyi model and the Generalized model but can be considered as a variation of the Generalized model. We impose certain constraints on the connection of links but do not impose a degree distribution. The constraints come in the form of specifying the desired average degree and the desired maximum degree of the network under construction. This is because the real power grid networks cannot have very high degrees and the average degree is usually between 2.5 to 3.5. 
Although real world power grid networks cannot be modeled as random networks, as mentioned in sections above, the purpose of this paper is simply to show that topological changes play a role in affecting the robustness of the network. If a set of additional geographical, voltage and other constraints are given, it is possible to create different feasible topologies using this set of constraints, with different degrees of robustness.

\subsection{Analysis of the generated networks and comparison of their robustness with the test cases}
\begin{figure}[h]
\centerline{\includegraphics[width = 3.5in, height =  1.5in]{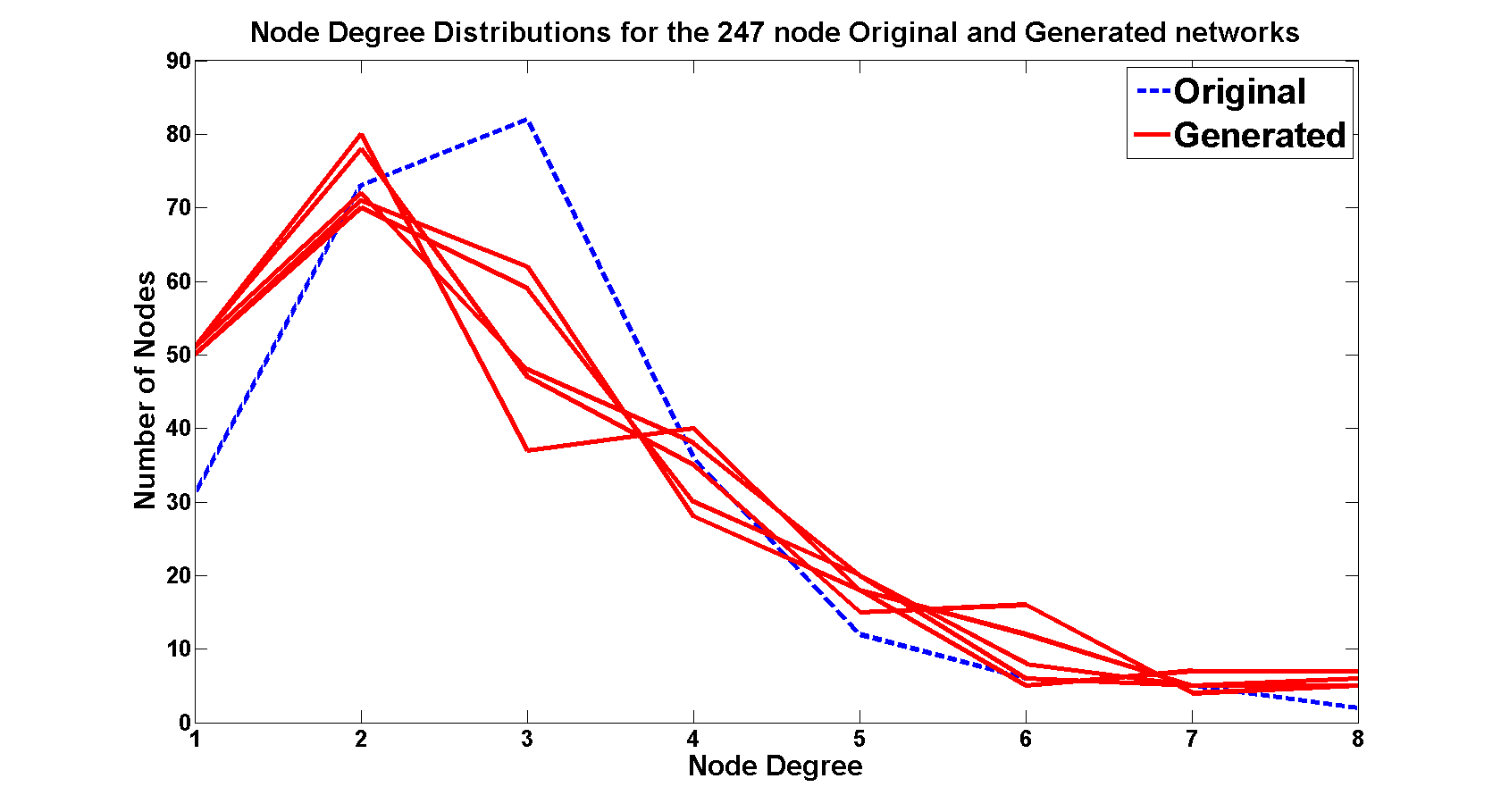}}
	\caption{Node Degree Distribution of the 247 Nodes Original and Generated Networks}
	\label{Fig4}
	\hfill
\end{figure}

\begin{figure}[h]
\centerline{\includegraphics[width = 3.5in, height = 2in]{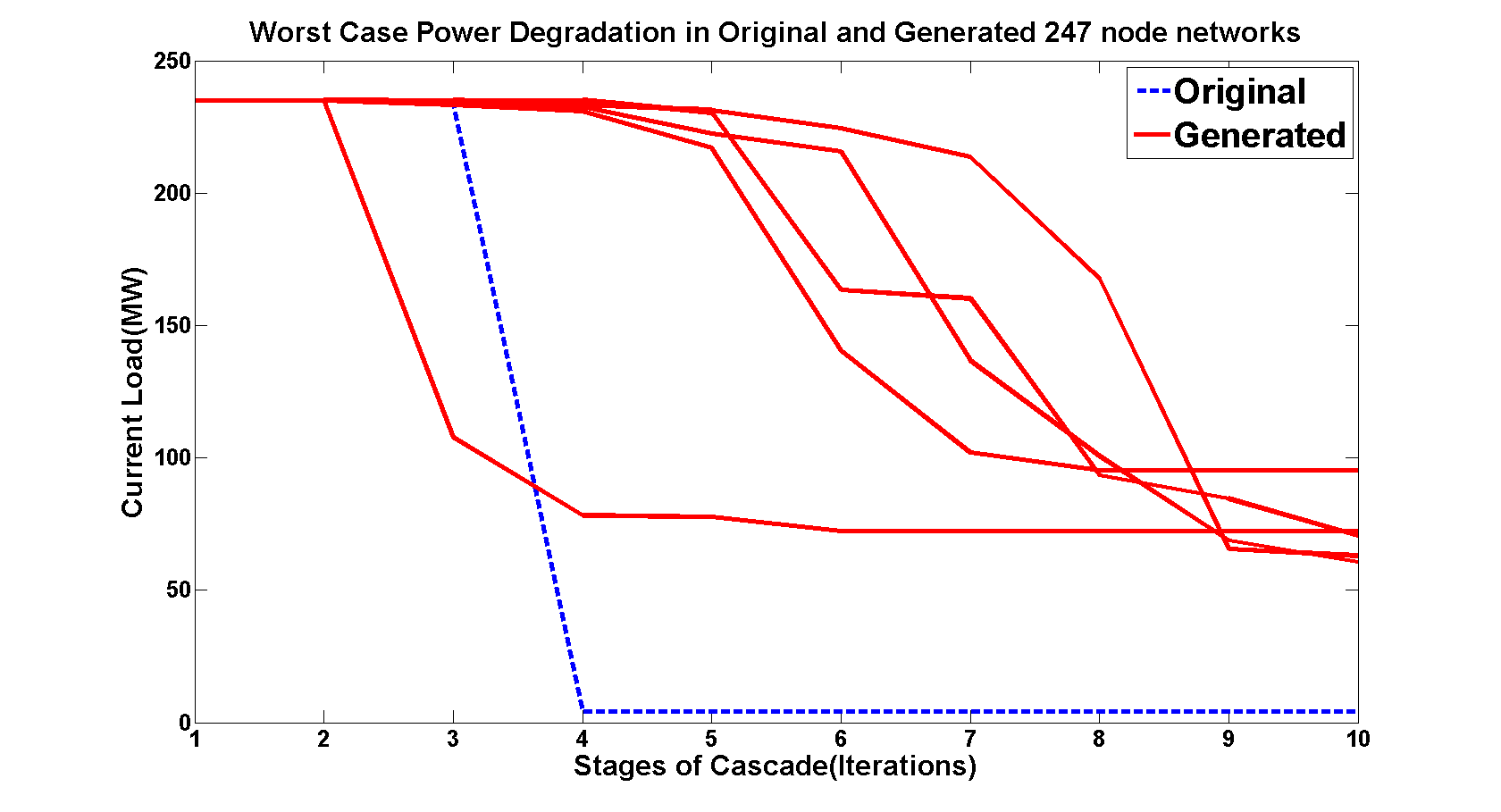}}
	\caption{Power Degradation Graphs for the 247 Nodes Original and Generated Networks}
	\label{Fig5}
	\hfill
\end{figure}

The generated random graphs closely follow the node degree distribution of the respective test case network, although it is not imposed. Fig. \ref{Fig1}, \ref{Fig2}, and \ref{Fig3} show the test network and five out of the family of generated networks for the 247 node, the 179 node, and the 118 node systems, respectively. Fig. \ref{Fig4} shows the node degree distributions of the 247 node test and generated networks. The horizontal scale represents the degree from $1$ to the maximum node degree for the networks (8 in this case), and the vertical axis represents the number of nodes with a given node degree. The generator takes the number of nodes, the maximum node degree, and the average node degree as inputs and gives the adjacency list of the network as the output. The adjacency list contains pairs of nodes that every link connects. Cascading failure is then simulated on these networks. We generate the impedances and loads probabilistically from the impedance and load distributions measured from the test networks. The adjacency matrix is created by random connection of nodes, taking into account that none of the nodes exceed the maximum node degree and there are no self loops or multiple edges between any two nodes in the system. 
Fig. \ref{Fig5} shows the graphs for power degradation on the 247 test and five generated networks. The horizontal scale represents the iterations that the system goes through before it finally stabilizes and the vertical scale is the current load on the system in MW. The generated networks have an average of 72.56MW of load remaining at stability but for the test network, less than 5MW remains. The original load on the system is 235.04MW. Each of the generated networks performs better than the test network but their behaviors differ from each other because each of them has a different topology and the efficiency of redistribution of power after the initial failure is different for different topologies. Table \ref{tab:Table1} shows the comparison between different topological characteristics of the test and the generated networks.

\begin{table}
		\begin{tabular}[width=0.5in]{|c|c|c|c|c|c|}
		\hline Nodes&Network&Path Length&Diam&Cluster Coeff\\
		\hline 
		\hline &$Test$ & $9.646$ & $24$ & $0.102$\\
					 &$G1$ & $5.162$ & $10$ & $0.0$\\
					 &$G2$ & $5.191$ & $11$ & $0.008$\\					 
					 $247$&$G3$ & $5.263$ & $13$ & $0.016$\\
					 &$G4$ & $5.190$ & $11$ & $0.006$\\
					 &$G5$ & $5.300$ & $12$ & $0.001$\\ 
		\hline &$Test$ & $12.382$ & $34$ & $0.089$\\
		       &$G1$ & $5.968$ & $14$ & $0.012$\\
		       &$G2$ & $6.058$ & $15$ & $0.0$\\
		       $179$&$G3$ & $5.661$ & $13$ & $0.012$\\
		       &$G4$ & $5.683$ & $13$ & $0.004$\\
		       &$G5$ & $5.616$ & $12$ & $0.001$\\
		\hline &$Test$ & $6.309$ & $9$ & $0.165$\\
					 &$G1$ & $4.223$ & $9$ & $0.032$\\
					 &$G2$ & $4.259$ & $9$ & $0.008$\\
					 $118$&$G3$ & $4.278$ & $9$ & $0.004$\\
					 &$G4$ & $4.348$ & $9$ & $0.004$\\
					 &$G5$ & $4.258$ & $9$ & $0.025$\\			 
		\hline
		\end{tabular}
	\caption{Differences in Characteristics of Test and Generated Networks}
	\label{tab:Table1}
\end{table}
The table discusses three network metrics: Characteristic Path Length, Diameter, and Clustering Coefficient.\\ Characteristic Path Length is the average shortest distance between all pair of nodes in the network. The shortest path between any given pair of nodes $i$ and $j$ has been measured in terms of number of links traversed by node $i$ to reach node $j$. The table indicates that the generated networks have a shorter characteristic path length as compared to the test networks. This is because random connections can lead to short cuts in the network with a given node connecting directly to a distant node. This makes the generated networks better connected and hence contribute to the improvement in robustness. Diameter is the longest shortest path among the set of shortest paths for all pair of nodes and is smaller for generated networks than that for the test networks. Shorter characteristic path length and shorter diameter ensure better global connectivity of the network. The Clustering Coefficient is a measure of the degree to which nodes in a graph tend to cluster. The local Clustering Coefficient of a given node $i$ is the degree to which its neighbors are connected to each other. The value of Clustering Coefficient shown in the table is the average value for the network. The Clustering Coefficient of the generated networks is much lower than that of the test networks due to the more random nature of the graph in the generated case. We are currently studying the contribution of clustering coefficient to the robustness of a power grid network and the drastic difference between the values for the test and generated networks indicates that it must have an effect on the robustness.
From a purely topological viewpoint, the generated networks are more robust than the test networks because of better connectivity.

\section{Mitigation Strategies}
In the past, some researchers have analyzed different mitigation strategies for blackouts in power grids. The authors of~\cite{BVID:02} have mentioned that their findings suggest, counter intuitively, that sometimes sensible attempts to mitigate failures in complex systems like power grids can have adverse effects and therefore must be approached with care. The authors have considered three types of mitigation measures and evaluated their impact on the frequency of large and small size blackouts. They suggest that mitigation measures, such as reducing the probability that an overloaded line suffers an outage, requiring a certain minimum number of lines to get overloaded before an outage occurs or increasing the generation margin shifts the dynamic equilibrium of the system and brings it to a point of self-organized criticality, reducing the risk of large blackouts. Similar analysis has been carried out in~\cite{JJI:05}. However,~\cite{PKE:09} has two other methods of mitigating cascading failures, namely survivability and reciprocal altruism. Other than these newly proposed strategies, load shedding has been a classical and a quite reliable mitigation strategy.
We propose the following three mitigation strategies, two of them based on load shedding, to limit the damage to the network by cascading failures: Homogeneous Load Reduction, Targeted Load Reduction and Islanding with distributed sources. Each of these strategies is discussed in detail.
\subsection{Homogeneous Load Reduction}
\begin{figure}[h]
	\includegraphics[width = 3.5in, height = 3in]{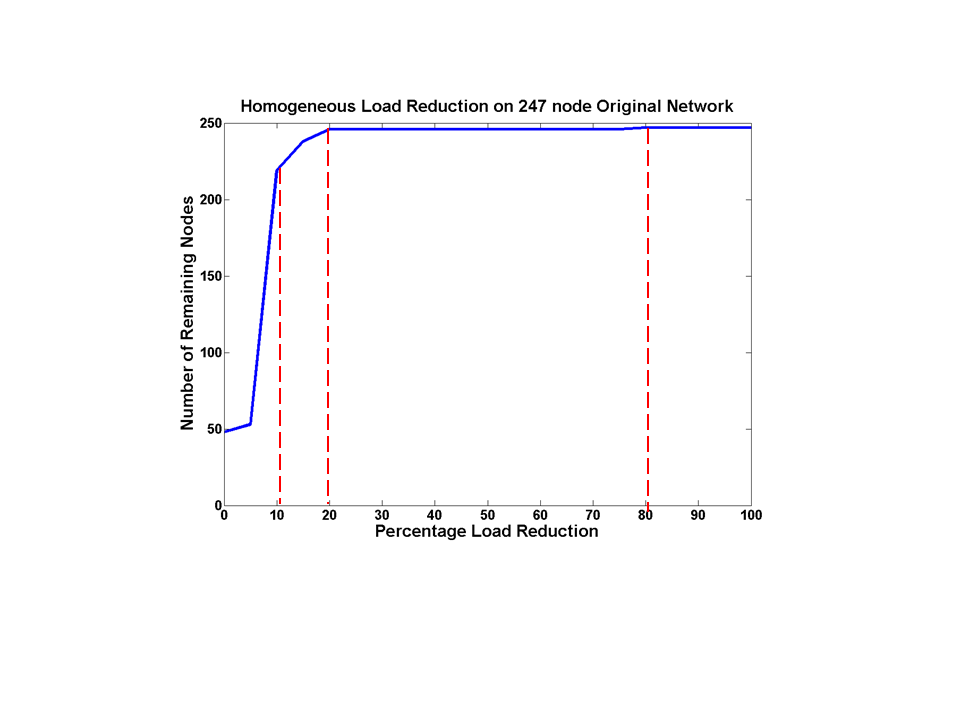}
	\caption{Homogeneous Load Reduction Strategy on the 247 Original network}
	\label{Fig6}
\end{figure}

This mitigation strategy aims at reducing a given percentage of the load on each of the nodes in the network. This reduction in load attempts to keep the nodes and links operating below their maximum capacities and to better accommodate the redistribution of power due to failure of links or nodes. We perform a series of simulations on the test 247 node network, wherein the load on each of the nodes is reduced from zero percent to hundred percent, in steps of five. Thus, the starting load for each simulation is the initial load on the nodes. We plot the result of each simulation for the test 247 node network to obtain the Homogeneous Load Reduction curve as shown in the Fig. \ref{Fig6}. Each point on the horizontal axis corresponds to one of the values of percentage reduction between zero and hundred. The graph represents the final result of the simulations corresponding to each point on the horizontal axis. About $10\%$ reduction on each of the nodes of the network leads to the emergence of a connected subgraph that contains the majority of the entire graph's nodes. However, the complete protection of the network takes place at $80\%$ load reduction, shown by the dotted line. Complete protection of the network means there is complete connectivity in the network or that all the nodes of the network are saved from failure. Yet there will be a very small amount of power degradation because of the initial failure, which is the first link that failed. The failure of the initial link does not necessarily lead to the failure of nodes.
\subsection{Targeted Load Reduction}
\begin{figure}
	\includegraphics[height = 2in]{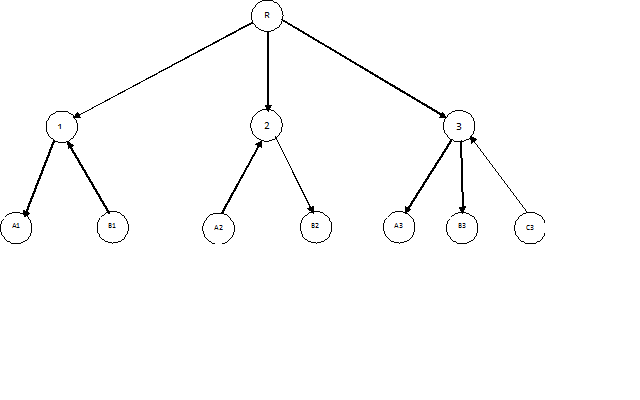}
	\caption{Load Reduction on each of the diamond-shaped nodes in the Tree - Targeted Load Reduction Strategy}
	\label{Fig7}
\end{figure}

The Homogeneous Load Reduction strategy is very simple to implement because it requires the same amount of reduction for each of the nodes in the network. But the drawback of this strategy is that load is reduced even from the nodes which may not be affected by re-routing of power to satisfy the demand. Thus, we propose a more efficient strategy, the Targeted Load Reduction strategy.
The objective of this strategy is to reduce the load only in a small portion of the network that will be affected due to the re-routing caused by the initial failure. If we select a node $i$ from the network and follow it along its adjacent nodes, considering the outgoing direction of power flow, we discover a tree, called the propagation tree, for node $i$. Fig. \ref{Fig7} would be helpful in understanding the approach. The approach used in this strategy is to select one of the node connected to the link that failed, and discover the tree for that node. It should be noted that out of the two nodes connected to the failed link, the one which was receiving power from that link should be used for tree discovery and load reduction because this reduces the demand and hence less re-routing is needed. 
The steps of the tree discovery algorithm for the Targeted Load Reduction strategy are listed below:

\begin{itemize}
\item Select a root node $R$.
\item Select the nodes directly connected to $R$ which receive power from $R$. These nodes are referred to as first-level nodes. Find the neighbors of the first-level nodes, other than $R$.
\item Select only those neighbors, other than $R$, of these first-level nodes that provide power to them.
\item For all the selected neighbors of these first-level nodes, including $R$, compare the magnitudes of the power supplied by each of them. The node which supplies highest power to a first-level node is on the maximum power flow path of that first-level node.\\
A maximum power flow path is the path which supplies the highest magnitude of power to a node from a generator, as compared to all the other paths. The back-tracing algorithm in Appendix $B$ describes the procedure to find the maximum power paths.
\item If $R$ is the node supplying maximum power to one or more of the first-level nodes, that first-level node becomes a part of the tree. Else that particular first-level node is discarded.
\item Now each of the first-level nodes selected in the tree act as a root node individually to select next possible candidates and the tree propagates.
\item The tree stops when we reach a source node or a leaf node. A leaf node is a node with degree one. 
\end{itemize}

After discovering the tree, an equal percentage of load is reduced on all the nodes of the tree except those nodes which do not have any load or which are sources. The analysis currently does not consider the priority of loads. There are some loads which must be supplied with power all the time and should not be considered for load reduction even if they are a part of the tree. However, all nodes are treated to be at the same priority in this work.\\
As shown in figure \ref{Fig7}, node $R$ is the initial root node. The node $R$ is connected to nodes 1, 2, and 3 referred to as the first-level nodes. Now we find the neighbors of these first-level nodes, other than $R$. Consider node 1 first. Node 1 has two other neighbors, $A1$ and $B1$. 

In this case, the root node $R$ and node $B1$ provide power to node 1 but root node $R$ supplies a higher magnitude of power to node 1 than node $B1$ does. Therefore, $R$ is on the maximum power flow path from the generator to node 1. This means that node 1 will be directly affected by the initial failure. Hence, node 1 forms a part of the tree and would be used for load reduction. Similarly, we find if node 2 and 3 form a part of the tree or not. 
The discovery procedure continues with node 1 as the new root node. Node 1 supplies power to node $A1$. In the next step, we consider the other neighbors of node $A1$ and find out if node 1 is the best supplier of power for node $A1$ or not. If it is, then node $A1$ forms the next level of the tree and the procedure continues for other nodes in a similar way. There may be some nodes in the propagation tree which may not carry any load but are simply used to transfer power from one node to the other. We reduce the load from the nodes in the tree, which carry load and are not source nodes. The tree terminates when it reaches a source node or a leaf node.

\begin{figure}
	\includegraphics[height = 2in]{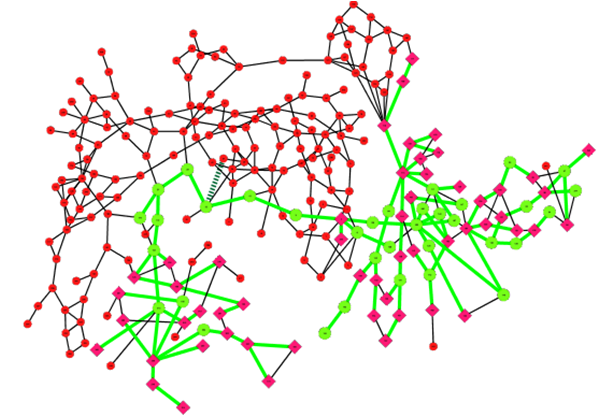}
	\caption{Tree view for the Targeted Mitigation Strategy}
	\label{Fig8}
\end{figure}

\begin{figure}
	\includegraphics[height = 2in]{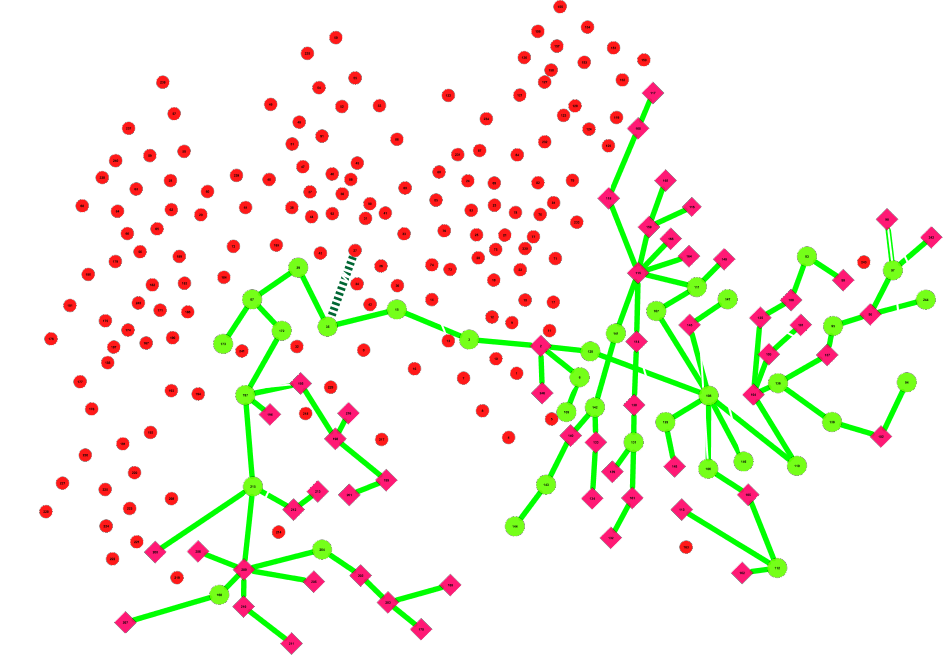}
	\caption{Targeted Mitigation Strategy showing only the tree and the disconnected link}
	\label{Fig9}
	\end{figure}

\begin{figure}
	\includegraphics[height = 2in]{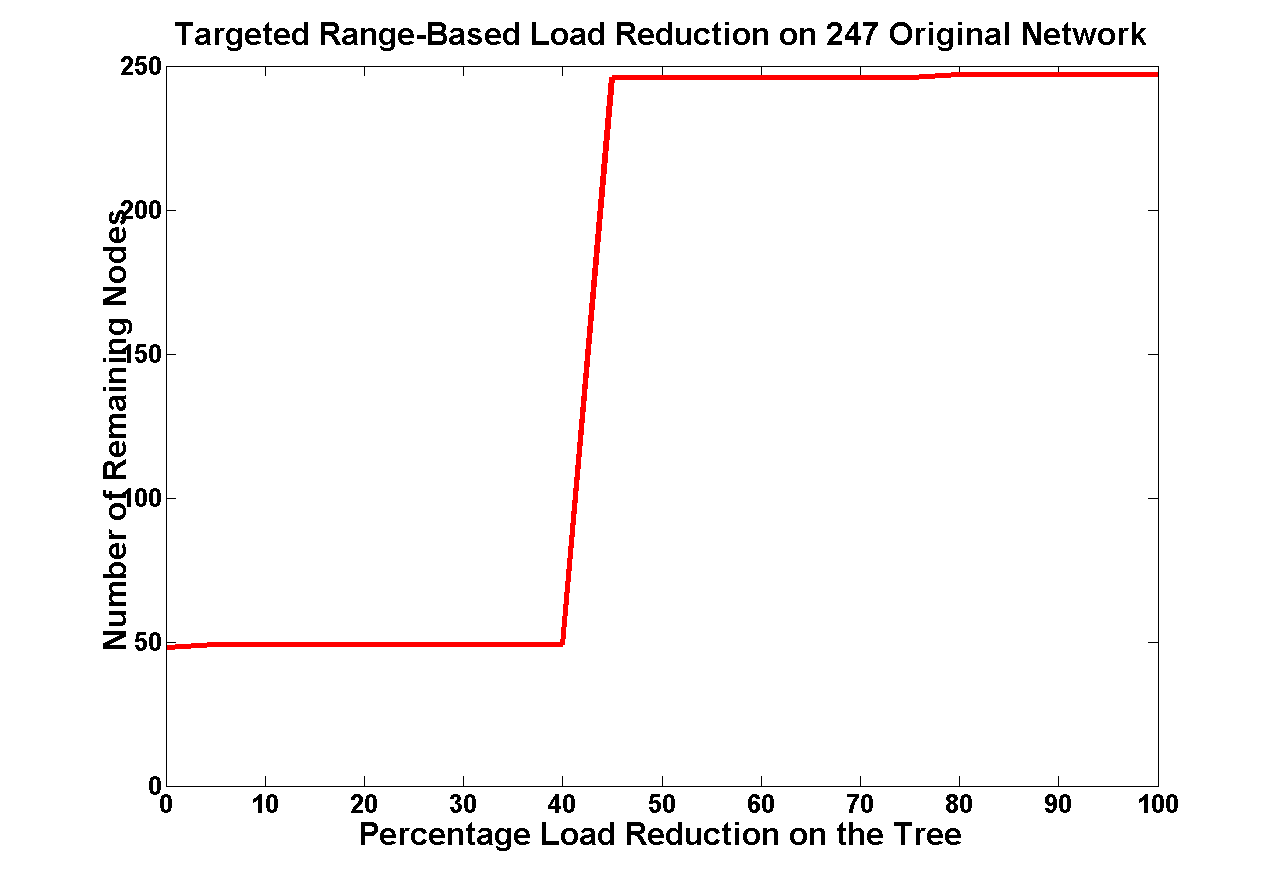}
	\caption{Targeted Mitigation Strategy - Load Reduction on the Tree}
	\label{Fig10}
\end{figure}

\begin{figure}
	\includegraphics[height = 2in]{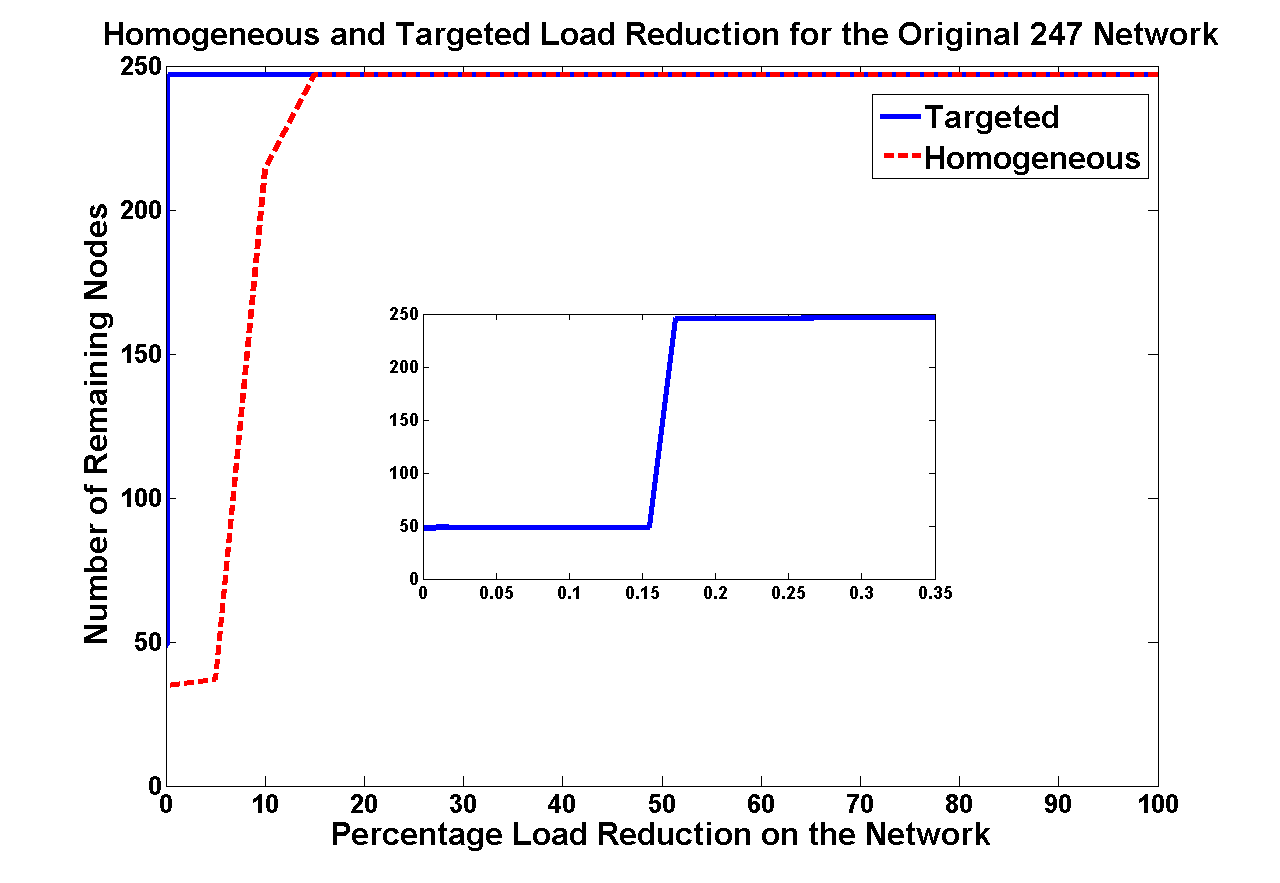}
	\caption{Homogeneous Load Reduction and Targeted Load Reduction on the network. Inset: Targeted Load Reduction on the network}
	\label{Fig11}
\end{figure}

The discovered tree for a given link in the 247 node test network is shown in Fig. \ref{Fig8} and more clearly in Fig. \ref{Fig9}. The total number of nodes in the tree is 90 but load is reduced on only 55 of these nodes. All the other nodes have no load on them. The magnified circular and diamond-shaped nodes form the tree, the diamond-shaped ones being used for load reduction. The thicker links are part of the tree and the broken link is shown dotted. Fig. \ref{Fig10} shows the graph of load reduction on the tree for a given failed link. A connected component with almost all the nodes is observed at $40\%$ reduction on the tree. The total protection of the network is achieved at $80\%$ reduction on the tree. But this constitutes a very small portion of the total load of the system, less than 2\%. Fig. \ref{Fig11} shows the comparison between the Homogeneous Load Reduction Strategy and the Targeted Load Reduction strategy for the given link, considering the entire network. The graph shows that the Homogeneous Load Reduction strategy helps to protect a major portion of the network at about $20\%$ load reduction. But the Targeted Load Reduction is more efficient because only about $0.35\%$ reduction on a small subset of the network is enough to get the same or better results as those obtained from the Homogeneous Load Reduction strategy, as shown in the inset.

Considering all the links in the network, the average size of the tree is $x$ nodes, with reduction happening on an average of $y$ nodes. The average reduction on the overall network is about $z$\%, to protect the network from a cascading failure.

\subsection{Distributed Generation and Islanding}

Distributed generation includes the application of small generators scattered throughout a power system to provide for the electric power needed by the consumers~\cite{HW:00}. In general, the term distributed generation refers to all the small electric power generators which are located on the utility system, at the site of a consumer. However, in this work, we deal with distributed generation at the transmission side to enable islanding of the transmission grid in the event of critical faults which may lead to a cascading failure. 

There are several partitioning algorithms discussed in literature for partitioning a network into "communities" or groups of nodes within which the connections are dense and between which they are sparse~\cite{MEJ:04}. Some of the graph partitioning computer algorithms are discussed in~\cite{BS:70}. However, as mentioned in~\cite{MEJ:04}, these algorithms are not ideally suited to general network analysis because they typically divide the network into two parts instead of a general number of communities and typically work for unweighted networks. An algorithm for community structure discovery in unweighted networks was proposed in~\cite{MM:02} to avoid the mentioned drawbacks of other algorithms. The same authors proposed a measure called "modularity", to evaluate the network partitions. Modularity is a quality function that determines the quality of the partitions on the scale of 0 to 1, with 1 being the highest value.

In this work, we create islands in the power grid network using a two-step process, followed by load shedding, if required. The first step gives a basic partitioning of the network using modularity~\cite{MEJ:04}, ~\cite{SF:10},~\cite{MN:04},~\cite{MN:06},~\cite{AC:05},~\cite{JPB:07},~\cite{MC:08},~\cite{AJAS:07},~\cite{EM:08},~\cite{YSH:10},~\cite{IBPS:10}.
The formula for modularity in weighted networks, as analyzed in~\cite{MEJ:04},~\cite{AJAS:07} and others is

\begin{equation}
Q = \frac{1}{2w} \sum_{i}\sum_{j}(w_{ij}-\frac{w_{i}w_{j}}{2w}) \delta(C_{i}, C_{j})
\end{equation}
where 
$w_{ij}$ is the weight on the link between node $i$ and node $j$ (0 if no link exists)\\
$w_{i}$ is the total weight on node $i$ \begin{equation} w_{i} = \sum_{j}{w_{ij}} \end{equation}\\
$w$ is the sum of the weights on all the links of the network \begin{equation} 2w = \sum_{i}{w_{i}} = \sum_{i}\sum_{j}{w_{ij}} \end{equation}\\
$C_{i}$ is the community to which node $i$ is assigned\\
and the Kronecker delta function $\delta(C_{i}, C_{j})$ is 1 if nodes $i$ and $j$ belong to the same community which means when $C_{i} = C_{j}$, else it is 0. 

The weights in this equation are replaced by power flows and the equation is modified as:

\begin{equation}
Q = \frac{1}{\Lambda} \sum_{i}\sum_{j}(P_{ij}-\frac{P_{i}P_{j}}{\Lambda}) \delta(T_{i}, T_{j})
\end{equation}
where
$P_{ij}$ is the power in the link between node $i$ and node $j$\\
$P_{i}$ is the total load on node $i$ \begin{equation} P_{i} = \sum_{j}{P_{ij}} \end{equation}\\
$\Lambda$ is the total load in the power grid network \begin{equation} \Lambda = \sum_{i}{P_{i}} = \sum_{i}\sum_{j}{P_{ij}}\end{equation}\\
$T_{i}$ is the island to which node $i$ is assigned\\
and the Kronecker delta function has a similar role as in the above definition of modularity.

\begin{figure}[h]
	\includegraphics[height = 2.5in]{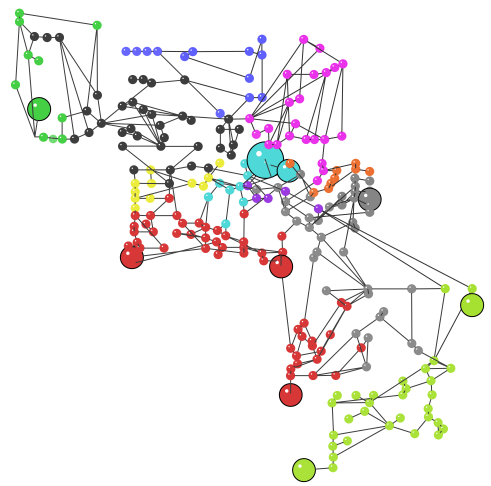}
	\caption{Islands in the 247 node network - Each color represents a different island with the big nodes being the sources. The biggest node is the main generator}
	\label{Fig12}
\end{figure}

\begin{figure}[h]
	\includegraphics[height = 2.5in]{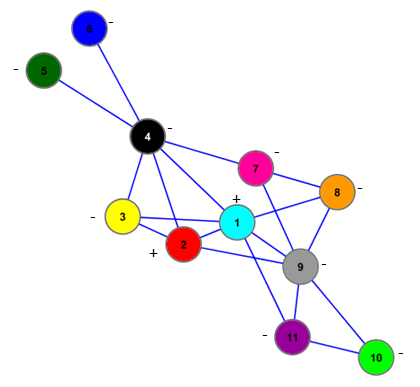}
	\caption{Aggregated network - Each island is represented as a single node. The islands with a '+' have excess generation while those with a '-' have less generation}
	\label{Fig13}
\end{figure}

This technique places low power flow links between islands and these links can then be disconnected without causing much damage to the network. The drawback of this technique is that it considers all nodes to be equal and does not distinguish between the distributed sources and other nodes. As a result, if the sources are not distributed evenly in space, one or more islands obtained using modularity may not have any generation. Thus, it is important to have a uniform spatial distribution of sources to obtain a good islanding scheme. However, if sources are not uniformly distributed, the drawback can be overcome by applying  the second step of the process, which is called 'Superislanding'. Superislanding means combining two or more islands such that the demand and supply within the superisland can be balanced with minimum load shedding. After the initial islanding, we convert the network to an aggregated network, with each island representing a single node in the system. Thus, the number of nodes in the aggregated network is equal to the number of islands obtained by modularity. We then create the adjacency matrix of the aggregated network for which the entry of the matrix is 1 if the islands are connected and 0 if they are not. The actual and aggregated network for the 247 node system is shown in Fig. \ref{Fig12} and Fig. \ref{Fig13} respectively. All nodes in the aggregated network have load while some nodes also have generation. The nodes with a '+' have more generation than required, and the nodes with a '-' have less generation than required, to fulfill the demand in that island. If a particular island has enough generation to satisfy its load, that island is not considered for superislanding.

The following algorithm describes the process of superislanding:
\subsubsection{Superislanding}
\begin{enumerate}
\item Assumptions:
\begin{enumerate}
\item The generation and the load are completely balanced in the initial system. In other words, total load in the system is equal to total generation.
\item No losses in the transmission line are considered.
\item Only those islands which have an imbalance in generation and load are considered for superislanding.
\end{enumerate}

\item Algorithm:
\begin{enumerate}
\item Check if island $i$ and $j$ are connected, $A(i,j) == 1$, and both are marked for superislanding.
\item Do a pairwise comparison of generation and load over all such connected pairs of nodes in the aggregated network:\\
$P_{g}(i,j) = P_{g}(i) + P_{g}(j)$\\
$P_{l}(i,j) = P_{l}(i) + P_{l}(j)$\\
$net(i,j) = P_{g}(i,j) - P_l(i,j)$\\
where, $P_{g}(i)$ is the individual generation of island $i$, $P_{g}(i,j)$ is the total generation of islands $i$ and $j$, $P_{l}(i)$ is the load of island $i$, $P_{l}(i,j)$ is the total load of islands $i$ and $j$, and $net(i,j)$ is the net generation of the combination. In case there are multiple combinations possible for a given island, we select the combination which gives the most positive $net(i,j)$ among the possible combinations of $i$ and $j$.
\item The following cases are considered for grouping islands:

$Case 1:$ Island $i$ has an excessive generation and island $j$ does not have any generation.\\
The most ideal case is that this combination does not make $net(i,j)$ negative, which means the total load does not exceed the total generation after combination. If $net(i,j) <$ 0 but total load exceeds the total generation by less than or equal to 5\%, excessive load is shed from $j$ and the combination is allowed.

$Case 2:$ If $i$ has excess generation and $j$ has a source but less generation. 
Again, the combination should not cause the condition $net(i,j) <$ 0 and if it happens the difference should be less than or equal to 5\% to allow superislanding.

$Case 3:$ If none of the above cases occur, load shedding is carried out for the island which has load imbalance. This makes the load and generation balanced and this island is no longer considered for superislanding. 

\item After all possible combinations, go back to step two and repeat the process until no further combinations are possible.
\end{enumerate}
\end{enumerate}

\section{Results}
The following results were obtained from the discussions above:
\begin{itemize}
	\item The power degradation graphs in Fig. \ref{Fig5} show that the worst-case cascading effect stops at an earlier stage and causes less damage in the case of generated networks. The load loss in the best topology of the 247 node generated network discussed above is about 36.79\% less than the test network.
	\item As seen from Table \ref{tab:Table1}, the generated graphs have a smaller characteristic path length and diameter as compared to the real networks. This property of the generated network topology makes them more robust against cascading failures. However, the clustering coefficient of the generated networks is much lower than that of the test networks, due to more random nature of the generated networks and its contribution to robustness must be analyzed.
	\item Targeted Load Reduction strategies form a special case of the Homogeneous Load Reduction strategy. The Homogeneous Strategy is easier to implement, in that we simply reduce the load on all the nodes existing in the network. The Targeted Load Reduction strategy is more economical, and reduces the load reduction to $x$\% as compared to the Homogeneous strategy, for an average of $y$\% network protection.\\
Formation of islands depends on the distribution of sources in the network. If the sources have an even spatial distribution, a good islanding scheme can be obtained. However, if the distribution of sources is not uniform, it leads to the formation of superislands.
\end{itemize}
\section{Conclusions and Future Work}
The topology of the power grid network contributes to its robustness. The topology determines the connectivity of the network and hence the number of alternate paths that can be taken by the network flow. The generated graphs are better connected because of shorter characteristic path lengths and are more random in nature than the real power grid networks. However, real power grid networks cannot be modeled in this way. Thus, if a set of geographical, voltage and other practical constraints is given, feasible power grid topologies can be created with different levels of vulnerability.

The Homogeneous Load Reduction strategy is similar to the classical mitigation strategies of load shedding and is simple to implement. The Targeted strategy is more efficient since the load is reduced on a very small subset of nodes and corresponds to a very small load reduction on the entire network.\\
Islanding incorporates the use of distributed renewable sources and helps to reduce the stress on the main grid by separating components from it without causing much damage and also ensures that the disconnected components are continuously powered without much loss of load.\\

Our future work includes the following:
\begin{itemize}
	\item Designing more feasible topologies taking into account geographical co-ordinates, 
	\item Study the impact of clustering coefficient on the robustness of the power grid network, and
	\item Thorough fault analysis with islanding and distributed sources
	
\end{itemize}
\section{Acknowledgments}
The authors would like to thank A. Pahwa, S. Starrett, I. Dobson, P. Schumm, and A. Hodges for their valuable suggestions and contributions to this work.

\begin{appendix}

The DC Power Flow model has been used for analysis of the topological features of the power grid network. For the DC model, we regard the branch resistances to be very small as compared to the inductive reactances and hence the resistances have been neglected. The shunt admittances have been neglected too. The line inductances govern the amount of power that flows through the links. The optimal path for power flow is the one that offers least impedance from a generator to a node or in other words, delivers the maximum power to a node from the generator. Node $1$ is considered as the reference node.
\subsection{DC Power Flow Model}
We cannot use metrics such as number of hops or shortest distance to find the efficient path from the generator to the destination node in case of a power grid. This inability to use such metrics arises from the fact that the amount of power flowing through a link depends upon the impedance of the link.\\
The map of the power grid in the form of nodes and links is used to create a weighted adjacency matrix $X$. When nodes $i$ and $j$ are connected, $x_{ij}$ is a positive non-zero entry and it is zero otherwise. These entries are the inductances of the transmission lines. A matrix $b$ is created from the adjacency matrix using the following formula~\cite{BR:88}:
\begin{equation}
b_{ij} = \frac{-1}{x_{ij}}, \forall i \neq j
\end{equation}
\begin{equation}
b_{ii} = \sum_{j=1}^{N}-b_{ij}
\end{equation}

The power handled by each node is the total load incident on the node. This is true because of the assumption that the total power going into the node is equal to the total power coming out of it.\\
Power in each link must be calculated as follows:

\begin{equation}
P_{ij} = -b_{ij}\delta_{ij}\\
			 = \frac{\delta_{i} - \delta_{j}}{X_{ij}}, \text{where $\delta_{ij}$ = $\delta_{i}$ - $\delta_{j}$}		 
\end{equation}

The power at each node can be calculated as:
\begin{equation}
P_{i} = \sum_{i=1, j\in n(i)}^{N}{P_{ij}} = \sum_{i=1, j\in n(i)}^{N}{-b_{ij}\delta_{ij}}
\end{equation}
where $n(i)$ = set of neighbors of $i$

The phase angles, $\delta$, are calculated for each node using the inverted $b$ matrix and a load vector $P$. This load vector is the load information that is given in the data sheet provided for the test networks. 

\begin{equation}
\delta = b^{-1}P
\end{equation}

We usually have the system ground is the reference bus in real power systems. The reference bus has the phase angle of $0\deg$ and all other angles are measured with respect to this angle~\cite{BR:88}. However, since we drop all shunt admittances for simplification, we lose the reference. This means that the $b$ matrix obtained from one of the above equation will be a singular matrix. We cannot calculate the inverse of the $b$ matrix if it is singular. To overcome this, one of the buses is assigned as the reference bus and the row and column corresponding to this reference bus in $\{b\}$ are dropped to produce a non-singular $\{b\}$ and its inverse is calculated. Following the real world system, we assign one of the generator nodes as the reference node and drop the row and column corresponding to it to produce a non-singular $\{b\}$. 

\subsection{Back-Tracing Algorithm to find the Maximum Power Paths}
The steps of the algorithm are described below. The algorithm back traces the path from a destination node to a generator. 
This path carries the maximum amount of power to the destination node out of all the paths that the node can trace back to other generators.
%Insert a sample 14 node picture here%

\begin{enumerate}
	\item Search for the neighbors of the selected destination node from the adjacency matrix. A given node is a neighbor if
	\begin{equation}
	b_{ij} \neq 0, \forall j \neq i
	\end{equation}
	\item Assume outgoing power to be positive and incoming power to be negative. Select the neighboring nodes which supply power to the destination node. In other words, the nodes providing power to the destination node are candidate previous nodes in the optimal path.\\
	\begin{equation}
	P_{ij} < 0, \forall j \neq i \Rightarrow 
	\end{equation} 
	$j$ is a possible candidate for the previous node in the path\\
	\item Compare the magnitudes of link power between the destination node and each candidate previous node and select the one with the highest magnitude.
	
	\begin{equation}
	\vert P_{ij} \vert > \vert P_{ik} \vert \Rightarrow
	\end{equation}
	previous node = $j$\\
	Else previous node = $k$ where $j$, $k$ $\in$ neighbors of $i$\\
	\item Now the previous node becomes the destination node and the same back tracing procedure is applied to search for its previous node and so on till we reach a generator.
	%Write about the figure that gives a visual representation of the algorithm here.%
\end{enumerate}
\end{appendix}

\bibliographystyle{IEEE}
\bibliography{PowerGrid_Cat}
\end{document}